\documentclass[aps,twocolumn,superscriptaddress,showkeys,groupedaddress]{revtex4}  % for review and submission
\usepackage{graphicx}  % needed for figures
\usepackage{dcolumn}   % needed for some tables
\usepackage{bm}        % for math
\usepackage{amssymb}   % for math

\begin{document}

\title{Direct Measurement of Turbulent Shear  }
%Note: in Physica D {\bf 240}, 1873 (2011)
%\affiliation{Department of Physics and Astronomy, University of Pittsburgh, Pittsburgh, Pennsylvania 15213, USA}
%\affiliation{Department of Physics, Ohio State University, Columbus, Ohio 43210, USA}

\author{S.~Stefanus} \thanks{ Corresponding author.\\ {\it E-mail address} : sts65@pitt.edu (S. Stefanus)\\ Phone: +1 (412) 624-9385} \affiliation{Department of Physics and Astronomy, University of Pittsburgh, Pittsburgh, Pennsylvania 15213, USA}

\author{S.~Steers} \affiliation{Department of Physics and Astronomy, University of Pittsburgh, Pittsburgh, Pennsylvania 15213, USA}\affiliation{Department of Physics, Ohio State University, Columbus, Ohio 43210, USA}
\author{W.I.~Goldburg} \affiliation{Department of Physics and Astronomy, University of Pittsburgh, Pittsburgh, Pennsylvania 15213, USA}

\vskip 0.2cm
\date{\today}

\begin{abstract}
A photon correlation method is introduced for measuring components of the shear rate tensor in a turbulent soap film. This new scheme, which is also applicable to three-dimensional flows, is shown to give the same results as Laser Doppler velocimetry, but with less statistical noise. The technique yields the mean shear rate ${\overline s}$, its standard deviation $\sigma$, and a simple mathematical transform of the probability density function $P(s)$ of the shear rate itself.
\end{abstract}


\maketitle

\section{Introduction}
Fluids dissipate energy as they flow through pipes or past any smooth or rough surface. Examples include river flow or wind blowing across the land. This energy dissipation is proportional to the velocity gradient, or shear rate, of the flow at the bounding surface. This frictional energy loss, and its dependence on the Reynolds number of the flow \cite{pope2000, tuan2010}, is not yet fully understood a century after the first explanation was advanced by L. Prandtl \cite{slichting2000}. 

Here we introduce a new scheme for measuring the shear rate near a bounding surface. It also might be applicable in the interior of a fluid. Unlike some widely-used methods \cite{pope2000}, the shear rate $s$ is recorded at a single "point" of size $w$. The motivation for developing this technique was to improve the usual method for measuring the shear rate in turbulent flows \cite{tuan2010,berman1973}.

The scheme introduced here is that of photon correlation spectroscopy (PCS) \cite{chu}. It is a variant of that used by Fuller and Leal to study laminar flows \cite{fuller1980}. For turbulence, the shear rate $s$ is a random variable. The PCS method enables determination of the time-averaged shear rate ${\overline s}$, its standard deviation $\sigma$, and the gaussian transform of the probability density function (PDF) $P(s)$ itself. Because the method has not been used before, the values of the mean shear ${\overline s}$ obtained by PCS are compared with those measured by laser Doppler velocimetry (LDV) \cite{drain}. 

The PCS scheme has the advantage of improved signal-to-noise, short data-collection times, and also the compactness of the apparatus. The PCS scheme can be used when the mean flow rate is absent or present. Hence it may be useful outside of the domain of turbulence studies. With the PCS scheme, a single beam illuminates a group of moving particles that scatter light into a photodetector at some scattering angle $\theta$. The inset of Fig.\ref{setup}a shows the incident and scattered laser beam of momentum ${\bf k}_0$ and ${\bf k}_s$, respectively and the scattering vector ${\bf k}= {\bf k}_s-{\bf k}_0.$ At a point in the flowing soap film, an incident beam is focused to a bright spot of size $w$. The intensity $I_0$ of the incident beam is taken to be gaussian form, $I_0(r)=I(0) e^{-(r^2/w^2)}.$ Figure \ref{setup}b, a side view of the setup, will be discussed below.  

The velocity at any point $r$ can then be written as the velocity at the center of the spot $r$=0 plus a term proportional to the shear rate tensor ${\tilde S}$, which is the quantity of interest. The dominant component of $\tilde S$ near a wall in this experiment is $s \equiv \partial_y u(y,t)$, where $u$ is in the flow direction $x$ while $y$ is in the transverse direction in the film plane.  Note that $s$ is a scalar quantity. Let $u(t)$ be the velocity of an illuminated particle at a horizontal distance $y$ from the center of the incident beam ($y=0$).  Then
\begin{equation}
u(y,t) = u(0,t) + s \,y(t) + ...,
\end{equation}
where the higher order terms have been neglected. 

Within a multiplicative constant, the scattered electric field from $N$ particles within the incident beam at time $t$ is
\begin{equation}
%scatteringfield eq 2
E(t)=\sum_{j=1}^N  E_0({\bf r}_j)e^{i{\bf k}\cdot {\bf r}_j(t)}
\propto \sum_{j=1}^N  E_0({\bf r}_j)e^{is({\bf k}\cdot {\bf r}_j) t}.
\end{equation}
Here $E_0({\bf r}_j)$ is the incident gaussian field at the position of the $j^{th}$ particle. 
Because the scattering from micron-size particles is almost perfectly elastic, $k=(4 \pi n/ \lambda)\sin (\theta/2)$ where $\lambda$ is the vacuum wavelength of the incident light beam (633 nm) and $n$ is the refractive index of the soap film, which is 99 \% water.

It will first be assumed that the flow is laminar, so that ${\tilde S}$ is time-independent, that is to say, the PDF of the shear tensor is a delta function centered at the mean value of ${\tilde S}$. Then the intensity correlation function which is simply related to the electric field autocorrelation function through the Bloch-Siegert theorem \cite{chu} (which is applicable to any gaussian PDF, including a delta function) is $g(\tau) \equiv \langle I(t)I(t+\tau) \rangle /\langle I (t) \rangle ^2=1+G(\tau),$ where
\begin{equation}
%blochsiegert eq 3
G(\tau)= |\langle E(t) E^*(t + \tau) \rangle|^2/ \langle I(t) \rangle ^2.
\end{equation}

Evaluating (3), using (1)-(2) and averaging over $t$ gives a result previously obtained by Fuller et al. \cite{fuller1980} for laminar flow, as opposed to a turbulent one. They evaluated $G(\tau)$ rather than $g(\tau)$. In the experiments described below, the turbulent soap film flows in the $x$ direction with mean velocity $U$, where this average is over the width $W$
of the soap film.  Then $G(\tau) = e^{- k^2 w^2 s_t^2 \tau^2/2},$
where $s_t$ is an average over time. Use has been made here of the gaussian form of the incident beam. 

Because $s$ is a random function for turbulent flows, an additional average over $s$ is needed, giving
\begin{equation}
%PDFofs equation 5
G(\tau) = \int e^{- k^2 s^2 w^2 \tau^2/2}P(s)ds,
\end{equation}
(with $P(s)$ having its maximum near ${\overline s}$); $G(\tau)$ is the gaussian transform of $P(s).$ 

Two important parameters, in addition to $w$, are $W$ and the Reynolds number, $Re=UW/\nu$, where $\nu$ is the kinematic viscosity of the soap solution.

If the supporting walls that bound the film flow are not smooth, their roughness $R^*$ is another important control parameter. As in three dimensions one expects \cite{goldenfeld2006,gioia2006,guttenberg2009} the dimensionless frictional drag $f \equiv\nu s/ U^2$ to be independent of $Re$ when it is sufficiently large. It now depends on the ratio $R^*/W$ \cite{pope2000}. At intermediate values of $Re$, experiment \cite{tuan2010} and theory \cite{guttenberg2009} support the result $f = C Re^{-1/2}$, where $C$ is just a number, and in 3D flows, $f = C Re^{-1/4}.$

If the shear tensor has more than one component \cite{fuller1980}
\begin{equation}
%PDFofs eq 6
G(\tau) = e^{-2 D k^2 \tau-U^2 \tau^2/w^2} \int e^{- ( {\tilde S} \cdot{\bf k} )^2 w^2\tau^2/2}P(s)ds,
\end{equation}
with ${\tilde S}_{ij} ={\tilde S}_{ji}$ when the fluid is incompressible, as in this experiment \cite{pope2000}. The factors to the left of the integral take into account the extraneous effects of particle diffusion and transit time broadening; they are discussed later in the text.

\section{The experiments}

\begin{figure}
\includegraphics[scale=0.2]{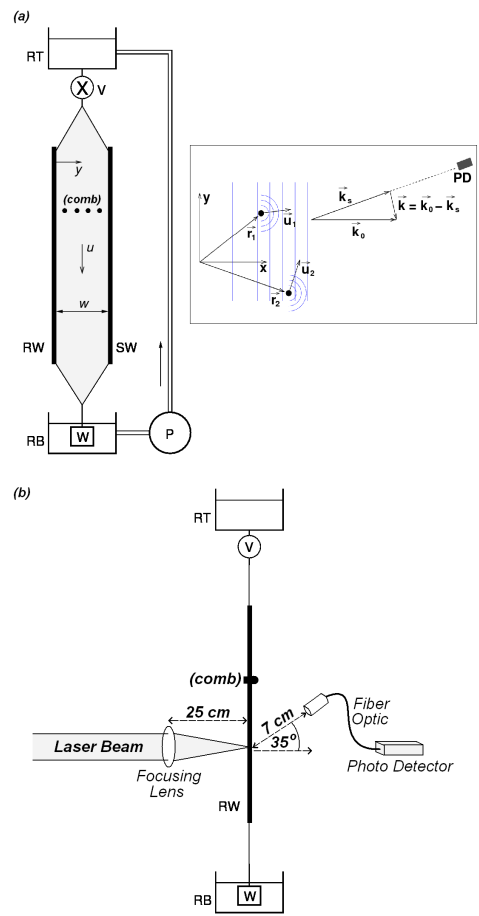}
\caption{\label{setup} (a): Setup for vertically flowing soap film. The film flows down from reservoir RT through valve V between strips RW and SW, separated by width $W$. The weight W keeps the the nylon wires taut. Inset shows scattering diagram. (b): Side view of the setup, showing laser source, focusing lens and photodetector}
\end{figure}

The experiments were performed on a soap film channel, shown in Fig.\ref{setup} with $W$=2 cm. The flow is driven by gravity, but there is an appreciable opposing force from air friction. However, near the vertical plastic strips that support the film, the viscous force from the wires dominates \cite{tuan2010}.

These strips are glued to thin plastic wires 0.5 mm in diameter that join to form an inverted V at the top and at the bottom, as indicated in Fig.\ref{setup}. At the apex, a small tube connects the reservoir to a valve V that controls the flow rate. The wires at the bottom connect and deliver the spent soap solution to reservoir RB, where it is pumped back to the top reservoir RT to keep the flow rate steady. Typical flow rates of the soap solution are $\sim$ 0.2 ml/s.

The soap solution is 1\% Dawn dishwashing detergent in water. It is loaded with neutrally buoyant polystyrene particles which scatter the incident beam from a 5 mW 633 nm He-Ne laser into the photodetector, a Perkin Elmer SPCM-AQR-12-FC. The laser source is located behind the soap film while the photodetector is located in front of it as shown in Fig.\ref{setup}b. The laser beam is focused onto the soap film with a lens of focal length 25 cm. The photon stream is delivered to the photodetector through an optical fiber, where the fiber tip is located 7 cm from the illuminated spot on the film. 

In these experiments, $w$ is limited by the wavelength of visible light, focal length of the focusing lens and the diameter of the incident beam and has a value of $w$ = 100 $\mu$m. The scattering vector ${\bf k}$ is in the vertical ($x$) direction and the dominant component of the shear rate is $s \equiv \partial_y u(y,t).$
The diameter $\phi$ of the seed particles (0.4 $\mu$m) is sufficiently small that their Stokes number in the strongest turbulence is less than 0.1 \cite{tuan2010}. Hence  the particle velocities are adequately close to that of the fluid. The refractive index of the soap solution is roughly 1.3. Typically, the scattering angle $\theta =35^\circ$, $k=6 \times 10^6$ m$^{-1}$. Using a seed-particle density of 1.5 gm/l yields an average photon counting rate of 10$^6$ Hz.

Experiments were performed with a horizontally oriented comb penetrating the soap film at a point above the measuring point and with the comb absent. Only with the comb present is the turbulence reasonably developed and the energy spectrum is of scaling form, $E(k) \propto k^{-b}$, with $b \simeq 3$ \cite{tuan2010,kraichnan,kellay2002}. This is the ${\it enstrophy \,range}$, defined as the interval where vorticity of larger size fluctuations cascade to smaller scales.  In two dimensions there is also a cascade of energy fluctuations to larger scales, where $b$ = 5/3, as in three dimensions. However, it is not accessible for decaying turbulence, as in this experiment \cite{kellay2002}. By making the bounding walls rough, so that turbulence is constantly being generated there, the inverse energy cascade can also be seen \cite{rutgers1998}. The teeth of the comb as well as their spacing is 2 mm.

To further test the PCS technique, measurements are also made with the comb absent. In this case, there is no well-defined energy spectrum decaying as a power law. Nevertheless the flow is far from laminar, so that ${\overline s}$ can be measured by both PCS and LDV.

%The most direct way to determine ${\overline s}$ by PCS is to measure $G(\tau)$ over a sufficiently small time interval that $k w \tau{\overline s}<<1$. In this limit, $e^{-k^2 w^2 s^2\tau^2/2}$ in (4) can be taken out of the integral, which is now unity with $s$ replaced by ${\overline s}$. So for small enough $\tau$, $G(\tau) = e^{-k^2 w^2 {\overline s}^2\tau^2/2}.$

\begin{figure*}
\includegraphics[width=3.3 in]{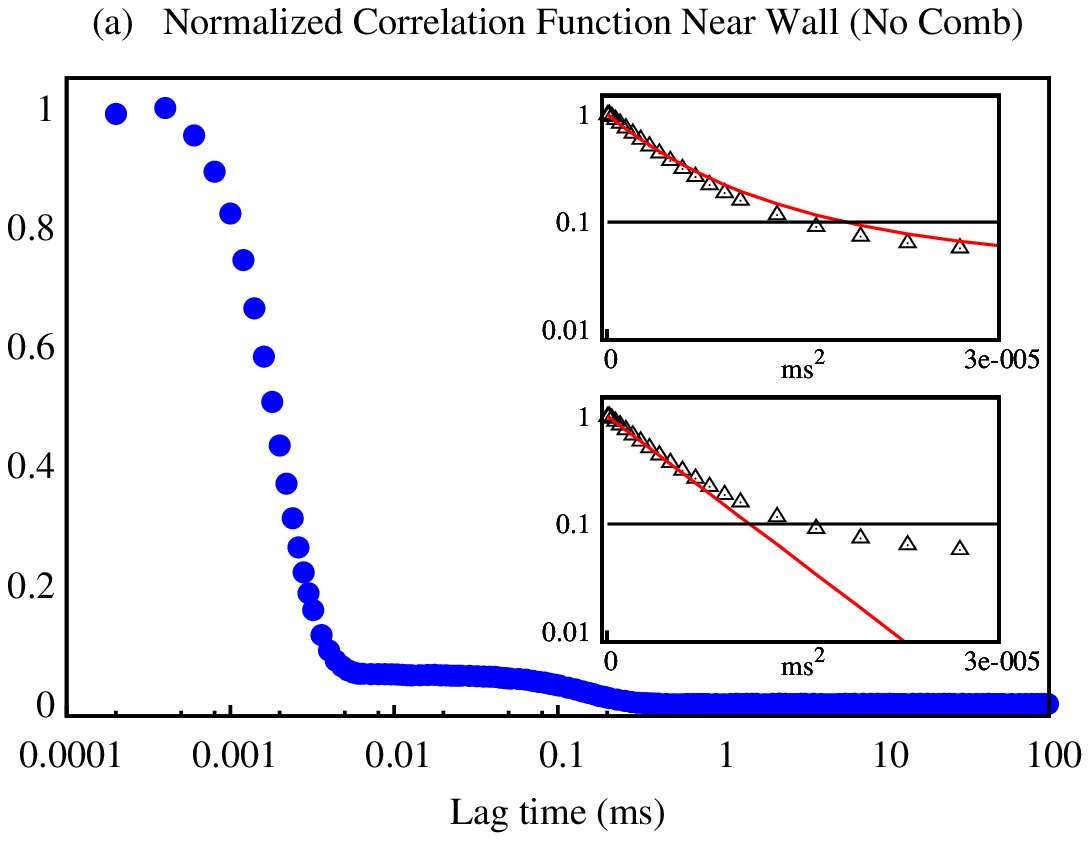}
\includegraphics[width=3.3 in]{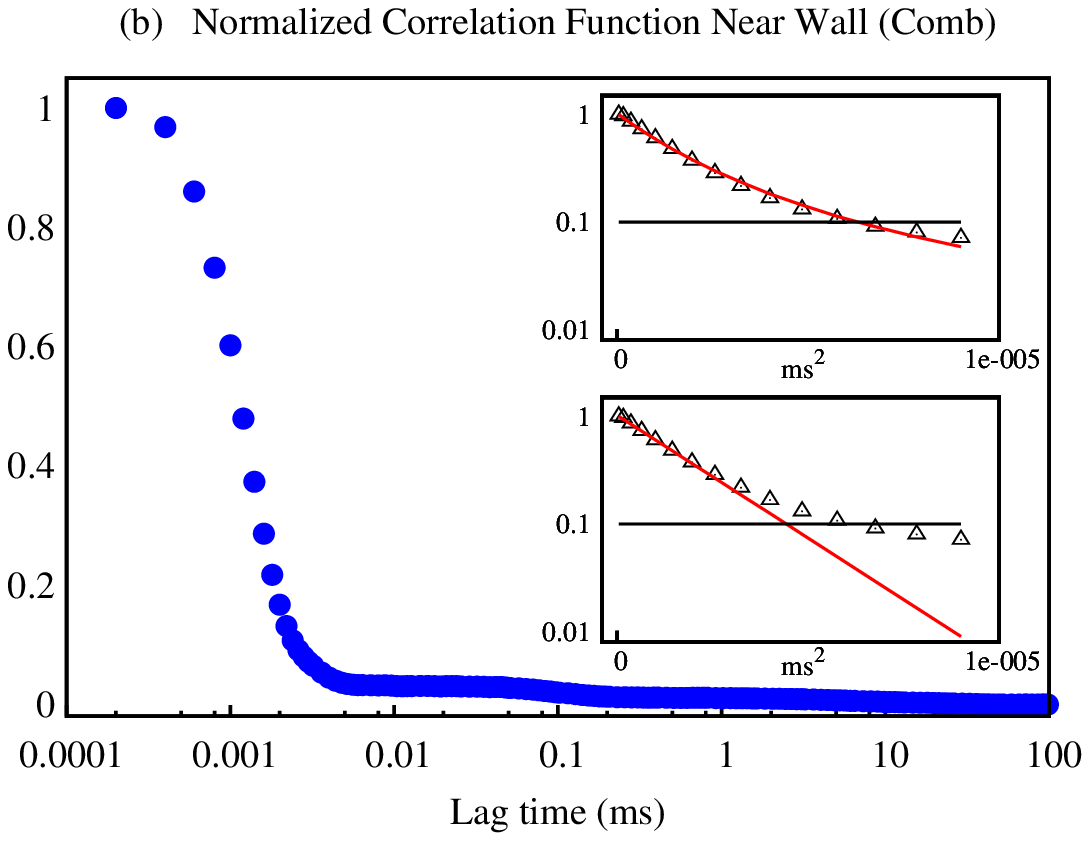}
\caption{Panel (a) and (b) are semilog plots of typical correlation functions $G(\tau)$ obtained with the comb absent and present. The conditions of these measurements are described in the text. The first data point in panel (a) should be ignored; it is instrumental in origin. The curved solid lines are a best fit to the data using a Gaussian $P(s)$. The straight solid lines are first order approximation of $G(\tau)$ and valid only for $k^2 w^2 {\overline {s^2}}\tau^2 << 1$. However, a Gaussian $P(s)$ fits $G(\tau)$ reasonably well for more than a decade. This fit has no theoretical basis.}
\label{corrWall}
\end{figure*}

To first order, $e^{-k^2 w^2 s^2\tau^2/2} \simeq 1 - k^2 w^2 s^2\tau^2/2$ and $G(\tau) \rightarrow 1 - k^2 w^2 {\overline {s^2}}\tau^2/2$. However, experimentally $G(\tau)$ is found to be a non-gaussian function.  If $P(s) \propto e^{-(s - {\overline s})^2/2 \sigma^2}$, 
%$G(\tau) = (k^2 w^2 \sigma^2\tau^2+1)^{-1/2}   e^{-k^2 w^2 {\overline s}^2 \tau^2/(2 k^2 w^2 \sigma^2\tau^2+2)}$, 
\begin{equation}
G(\tau) = \frac {1} {\sqrt{k^2 w^2 \sigma^2\tau^2+1}}   e^{-k^2 w^2 {\overline {s^2}} \tau^2/(2 k^2 w^2 \sigma^2\tau^2+2)},
\end{equation}
which is clearly non-gaussian in $\tau$. Both panels of Fig. \ref{corrWall} show that while $P(s)$ is close to gaussian form, the gaussian fit (solid lines) is not perfect.  These "good" fits to gaussian form were unexpected. 

There are two other effects that can contribute to the decay of $G(\tau)$: thermal diffusion of the seed particles and transit time broadening, which can be dominant for large $U/w$. Both of these contributions are small in these experiments but are easy to correct for \cite{chowdhury1984}. To take diffusion into account, one multiplies Eq. (4) by the factor,
$G_D =e^{-2Dk^2 \tau},$ where $D$ is the diffusion constant, which, for spherical particles of diameter $\phi$ is $k_BT/3 \pi \eta \phi$, where $k_B$ is Boltzmann's constant and $\eta$ is viscosity.

As for the transit time effect, particles passing through a beam of size $w$ produce a burst of light intensity that temporally modulates the scattered light. The multiplicative correction factor here is $G_{tt}(\tau) = e^{-(U^2\tau^2/w^2) }$\cite{chowdhury1984}.

The decay times for both of these effects is long compared to the viscous decay time of interest so these multiplicative time factors can be dropped. For example, with a spot size $w= 100 \mu$m and a typical mean velocity of $U$ = 1 m/sec, the transit time $\tau_{tt}$ associated with the effect is or order $a/U \simeq $ 0.1 ms. This is fifty times longer than typically measured $\tau_c$. Diffusion times are much longer than this and hence contribute insignificantly to the decay of $G(\tau)$.

Fig.\ref{corrWall} shows $G(\tau)$ for measurements made with the comb absent (a) and present (b), respectively. Here $U \simeq$ 2 m/s in both experiments, $W$=2 cm, and $w$ = 100 $\mu$m. The vertical axis is linear, but the horizontal axis is $\log \tau$, so as to display several decades of lag time. The insets to both figures show $ \log G(\tau)$ vs $\tau^2$, so that a gaussian decay of $G(\tau)$ appears as a straight line. The straight lines in the lower insets indicate that $G(\tau)$ is indeed of gaussian form for very small $\tau$. They are a best fit to the experimental curves and correspond ${\overline s}$ = 1600 s$^{-1}$ and 1000 s$^{-1}$ for the experiments with and without the comb.

The solid lines in the upper insets to Fig.\ref{corrWall} are best fits under the assumption of a gaussian $P(s).$ A good fit clearly extends beyond the small-$\tau$ limit and enables the determination of the standard deviations $\sigma$ of the mean shear as well as ${\overline s}$ itself. The mean shear ${\overline s}$ is calculated from the definition of variance, $\sigma^2 = \overline{s^2} - {\overline s}^2, \overline s = \sqrt{\overline{s^2} - \sigma^2}$. The results are ${\overline s}$ = 950 Hz $\sigma$ = 300 Hz with the comb absent and ${\overline s}$ = 1620 Hz $\sigma$ = 500 Hz with the comb present. The ratio of $\sigma$ to ${\overline s}$ is near 20 \%.

The shear measurement is done in the viscosity-dominated layer of width $\delta(x)$, where $x$ is the distance from the comb. Ideally the spot size $w$ should be much smaller than $\delta(x)$. The function $u(y)$ is proportional to $y$ within $\delta(x)$. Prior experiments have established that at $x$ = 20 cm below the comb, where the measurements were made, $\delta$ is roughly 200 $\mu$m \cite{tuan2010}. Thus the beam size $w$ is small enough to correctly measure the viscous shear rate.

The single-point PCS measurements of ${\overline s}$ are now compared with those of LDV, made in the traditional way; the vertical velocity component $u$ is measured at two nearby horizontally-spaced points in the viscosity-dominated interval. 

The LDV measurements were made 2.5 cm below the PCS beam spot, which is 80 cm below point P in Fig.\ref{setup} and 20 cm below the comb. The LDV measurement point is advanced in 50 $\mu$m steps starting at $y$=0. The minimum useful value of $y$ is dictated by the necessity of avoiding strong light scattering from the supporting plastic strip with its edge at $y$=0.

The LDV laser source is 514 nm line from a Coherent argon-ion laser operating at a power of 500 mW, roughly one hundred times that used in the PCS device. The data collection time for each measurement of $u(y)$ is roughly 20 s. Because the correlation time is of the order of microseconds, and the counting rate is of the order of MHz the function form of $G(\tau)$ emerges after only a few seconds of data collection with the correlator.

\section{Results}

The mean shear rate ${\overline s}$ in the viscous region obtained by LDV and PCS agree to within one standard deviation, as seen in Table \ref{tab:comb_shear_rate}. The uncertainties are deduced from seven measurements made at the indicated values of $U$. From an individual run, one cannot extract $\sigma$ from the LDV data, because noise fluctuations can change even the sign of $\partial_y u(y,t)$.

The LDV and PCS measurements span the range 29000 $< Re < $45000 and from 40000 $< Re < $57000, with and without comb respectively.   With the comb in place, the Taylor microscale Reynolds number $Re_\lambda \equiv u_{rms} \lambda/\nu = 130$, where $\lambda =u_{rms} /\sqrt{\langle (\partial u/\partial x)^2\rangle}$ = 1 mm.

The errors from one run to another are {\it not} statistical in origin. Rather, the source is variations in the flow speed through the valve and the motion of the film plane caused by velocity fluctuations of the surrounding air which could be only partially suppressed by placing the entire apparatus in a tent.

\begin{table}
\caption{\label{tab:comb_shear_rate}Mean shear rate ${\overline s}$ as measured by LDV and PCS in a narrow range of mean flow speeds (comb inserted).}
\begin{ruledtabular}
\begin{tabular}{ccc}
{\bf $U$ (m/s)} & LDV ${\overline s}$ (Hz) & PCS ${\overline s}$ (Hz)  \\
\hline
1.4 & 1500& 1660 \\
1.6 & 1120 & 1030 \\
1.7 & 1760 & 1830 \\
1.8 & 1880 & 1430 \\
1.9 & 1230 & 1200 \\
2.2 & 1760& 1700 \\
2.2 & 1860 & 1640 \\
\hline
{\bf Average}  & 1590 & 1500 \\
{\bf Std. Dev.} & 300 (20\%) & 300 (20\%) \\
\end{tabular}
\end{ruledtabular}
\end{table}

\begin{figure}
\includegraphics [width=3 in]{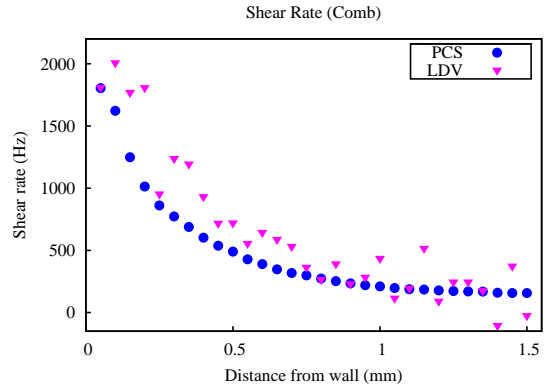}
\caption{Plot of mean shear rate ${\overline s}$ as a function of distance from the wall (in mm) with a comb in place to strengthen the turbulence. The mean flow speed $U$ = 2.16 m/s, $Re$=45,000.} \label{PCS_LDV_Comp_15}
\end{figure}

Fig.\ref{PCS_LDV_Comp_15} shows measurements of ${\overline s}$ as a function of $y$ in units of 50 $\mu$m obtained using PCS (circles) and LDV (triangles) in the range out to $y$ =1.50 mm with the comb present. Here $U$ = 2.16 m/s, $W$=2 cm and the kinematic viscosity of the soap solution is close to that of water ($\nu$ =0.01 cm$^2$/s), $Re$=45,000. 

The main messages conveyed by this graph are (a) the two schemes give roughly the same results for ${\overline s}(y)$, (b) near one of the walls, the LDV measurements are noisier (for reasons already discussed), and (c) ${\overline s}$ decreases with increasing $y$. Even in the absence of air friction, this decrease is expected and is well-studied in 3D flows \cite{pope2000}. In soap film flows, air friction slows the flow far from the walls, making analysis of the data there difficult. 

This experiment indicates it should be possible to measure ${\overline s}$ near the wall and in the interior of 3D flows, though care must be taken to collect scattered photons from only a small volume in the fluid. Far from a bounding wall in 3D turbulence, the PCS method will suffer from the limitation that $w$ should be smaller than the smallest eddy size \cite{frisch}, defined as $\eta =(\nu^3/\epsilon)^{1/4}$.  Even in these soap film experiments, $\eta$ is estimated to be comparable to or smaller than $w$.  Yet, as Fig.\ref{PCS_LDV_Comp_15} shows, the LDV measurements of ${\overline s}$ agree with the PCS result up to $y$ = 1.5 mm from a wall, well outside the viscous region.

\section{Summary}
Though the photon correlation scheme has been used here to measure properties of the shear rate in a two-dimensional soap film, it can be used in three dimensional flows as well. The PCS method has good signal-to-noise, is compact, and uses a laser in the mW range. The method yields the variance of the shear rate as well as its mean value. The correlation function itself is the gaussian transform of the probability density function, $P(s)$.

\section{Acknowledgments}
We wish to thank T. Tran, T. Adamo, W. Troy, N. Goldenfeld, N. Guttenberg, and K. Nguyen for their contributions to this work. This work is supported by NSF grant No. DMR  0604477 and NSF Fellowship 60016281 to S. Steers.

\end{document}